\documentclass[%
 reprint,
showpacs,
 amsmath,amssymb,
 aps,
 prl
]{revtex4-1}

\usepackage{graphicx}
\usepackage{dcolumn}
\usepackage{bm}
\usepackage{color}

\begin{document}

\preprint{APS/123-QED}

\title{Identification of Spin-Triplet Superconductivity \\ through a Helical-Chiral Phase Transition in Sr$_2$RuO$_4$ Thin-Films}

\author{S. Ikegaya$^{1}$, K. Yada$^{2}$, Y. Tanaka$^{2}$, S. Kashiwaya$^{2}$, Y. Asano$^{3}$, D. Manske$^{1}$}
\affiliation{$^{1}$Max-Planck-Institut f\"ur Festk\"orperforschung, Heisenbergstrasse 1, D-70569 Stuttgart, Germany\\
$^{2}$Department of Applied Physics, Nagoya University, Nagoya 464-8603, Japan\\
$^{3}$Department of Applied Physics, Hokkaido University, Sapporo 060-8628, Japan}

\date{\today}

\begin{abstract}
Despite much effort for over the two decades, the paring symmetry of a  Sr$_2$RuO$_4$ superconductor has been still unclear.
In this Rapid Communication, motivated by the recent rapid progress in fabrication techniques for Sr$_2$RuO$_4$ thin-films,
we propose a promising strategy for identifying the spin-triplet superconductivity in the thin-film geometry
by employing an antisymmetric spin-orbit coupling potential and a Zeeman potential due to an external magnetic field.
We demonstrate that a spin-triplet superconducting thin-film undergoes a phase transition from a helical state to a chiral state by increasing the applied magnetic field.
This phase transition is accompanied by a drastic change in the property of surface Andreev bound states.
As a consequence, the helical-chiral phase transition, which is unique to the spin-triplet superconductors,
can be detected through a sudden change in a tunneling conductance spectrum of a normal-metal/superconductor junction.
Importantly, our proposal is constructed by combining fundamental and rigid concepts regarding physics of spin-triplet superconductivity.
\end{abstract}

\maketitle

\textit{Introduction}---%
Since 1994, a great attention has been drawn to a Sr$_2$RuO$_4$ because it is a leading candidate material for spin-triplet superconductors~\cite{maeno_94,maeno_03,maeno_12}.
Actually, on the basis of a number of experiments~\cite{duffy_00, ishida_01, mukuda_02, murakawa_04, murakawa_07, jang_11, kashiwaya_11}
and theories~\cite{rice_95, nomura_00, nomura_02, yanase_03},
a Sr$_2$RuO$_4$ has been believed to exhibit spin-triplet chiral $p$-wave superconductivity with broken time-reversal symmetry~\cite{luke_98, xia_06}.
However, very recent experiments of nuclear magnetic resonance Knight shift at oxide sites~\cite{brown_19, ishida_19} seem to be inconsistent with this scenario,
and rather suggest the realization of spin-singlet superconductivity in this compound~\cite{simon_19}.
Such stalemate situation requires us to propose a refreshing experiment for identifying the spin-triplet superconductivity.
In this Rapid Communication, we show that the recent rapid development in fabrication techniques for Sr$_2$RuO$_4$ thin-films enables us to shed some light on this issue.

An essential character of a spin-triplet superconductor is that its order parameter is described by a three components vector, $\boldsymbol{d}$-vector,
reflecting a spin-degree of freedom in spin-triplet Cooper pairs.
Therefore, in essence, evidences of spin-triplet superconductivity are provided from observations of unique phenomena in the presence of the $\boldsymbol{d}$-vector.
A primary factor interacting with the $\boldsymbol{d}$-vector is magnetic potentials, such as Zeeman potentials and exchange potentials.
Accordingly, previous studies so far have mainly focused on phenomena of spin-triplet superconductors in the presence of magnetism,
such as the temperature-independent spin susceptibility~\cite{duffy_00, ishida_01, murakawa_04} and
the long-range proximity effect in ferromagnet/superconductor junctions~\cite{yoshida_99, hirai_01, boris_06, anwar_16}.
However, at present, the conclusive experimental evidence for the spin-triplet superconductivity has not yet been observed.
A substantial progress in this research field arises from the fabrication techniques for Sr$_2$RuO$_4$ thin-films~\cite{uchida_16, blamire_16, uchida_17, schlom_18, uchida_19, schlom_19}.
This movement enables us to employ a refreshing factor for the identification of the spin-triplet superconductivity:
antisymmetric spin-orbit coupling (ASOC) potentials due to broken inversion symmetry.

In this Rapid Communication, we study the gap function and transport property of a spin-triplet superconducting thin-film coexistence
with both an ASOC potential and a Zeeman potential due to an external in-plane magnetic field.
We first demonstrate that the spin-triplet superconducting thin-film can show
a phase transition from helical to chiral spin-triplet superconducting states by increasing the magnetic field.
This phase transition is essentially due to the characteristic nature of the $\boldsymbol{d}$-vector, and is never expected in spin-singlet superconductors.
Then, we show that this phase transition is detected through a drastic change in the tunneling conductance spectrum of normal-metal/superconductor junctions.
As we discuss later, our proposal is constructed by a combination of general and rigid concepts regarding physics of spin-triplet superconductivity.
Consequently, we indicate a promising strategy for identifying the spin-triplet superconductivity in Sr$_2$RuO$_4$ thin-films.

\textit{Helical-Chiral phase transition}---%
In this Rapid Communication, for simplicity, we focus only on the $\gamma$-band of Sr$_2$RuO$_4$
which is considered to play the dominant role for the superconductivity~\cite{Mackenzie_03, deguchi_06}.
We describe the superconducting states by the following two-dimensional single-band mean-field Hamiltonian,
\begin{align}
H &= \sum_{\boldsymbol{k}} \sum_{\alpha,\beta}
\left( \xi_{\boldsymbol{k}} \hat{\sigma}_0 + \boldsymbol{g}_{\boldsymbol{k}} \cdot \hat{\boldsymbol{\sigma}} + \boldsymbol{V} \cdot \hat{\boldsymbol{\sigma}} \right)_{\alpha\beta}
c^{\dagger}_{\boldsymbol{k}\alpha} c_{\boldsymbol{k}\beta} \nonumber\\
&+\frac{1}{2}\sum_{\boldsymbol{k}} \sum_{\alpha,\beta}
\left[ \Delta_{\boldsymbol{k},\alpha\beta} c^{\dagger}_{\boldsymbol{k}\alpha} c^{\dagger}_{-\boldsymbol{k}\beta} + \mathrm{h.c.} \right] \nonumber\\
&-\frac{1}{2}\sum_{\boldsymbol{k}} \sum_{\alpha,\beta} \Delta_{\boldsymbol{k},\alpha\beta}
\langle c^{\dagger}_{\boldsymbol{k}\alpha} c^{\dagger}_{-\boldsymbol{k}\beta} \rangle,
\end{align}
where $c^{\dagger}_{\boldsymbol{k}\alpha}$ ($c_{\boldsymbol{k}\alpha}$) is creation (annihilation) operator of an electron
with momentum $\boldsymbol{k}$ and spin $\alpha$,
the Pauli matrices in spin space are given by $\hat{\boldsymbol{\sigma}}=(\hat{\sigma}_x,\hat{\sigma}_y,\hat{\sigma}_z)$,
and the $2 \times 2$ unit matrix is denoted by $\hat{\sigma}_0$.
The kinetic energy of an electron measured from the chemical potential $\mu$ is given by
\begin{align}
\xi_{\boldsymbol{k}}=-2t(\cos k_x + \cos k_y) - 4t^{\prime} \cos k_x \cos k_y - \mu \nonumber
\end{align}
with $t$ and $t^{\prime}$ representing the nearest-neighbor and next-nearest-neighbor hopping integral, respectively.
To reproduce the Fermi surface of the $\gamma$-band, in what follows, we set $t=1.0$, $t^{\prime}=0.395$ and $\mu = 1.5$~\cite{nomura_00, nomura_02}. 
The ASOC potential is described by the $\boldsymbol{g}$-vector, $\boldsymbol{g}_{\boldsymbol{k}}=-\boldsymbol{g}_{-\boldsymbol{k}}$.
Although the momentum dependence of the $\boldsymbol{g}$-vector in real systems may be more complicated,  for simplicity,
we use the conventional Rashba type spin-orbit coupling potential $\boldsymbol{g}_{\boldsymbol{k}}=\lambda (\sin k_y, -\sin k_x, 0)$.
Even so, as we discuss later, the validity of our proposal is insensitive to the detailed structure of $\boldsymbol{g}_{\boldsymbol{k}}$.
The Zeeman potential due to an externally applied in-plane magnetic field is $\boldsymbol{V}=(V_x, V_y, 0)$.
The pair potential is represented by $\Delta_{\boldsymbol{k},\alpha\beta}$.
Within the weak-coupling mean-field theory, the pair potential is determined by the gap equation
\begin{align}
\Delta_{\boldsymbol{k},\alpha\beta}= \sum_{\boldsymbol{k}^{\prime}} \sum_{\gamma, \delta}
g_{\alpha \beta \gamma \delta}(\boldsymbol{k}, \boldsymbol{k}^{\prime})
\langle c_{-\boldsymbol{k}^{\prime}\gamma} c_{\boldsymbol{k}^{\prime}\delta} \rangle,
\label{eq:gap_eq}
\end{align}
where $g_{\alpha \beta \gamma \delta}(\boldsymbol{k}, \boldsymbol{k}^{\prime})$ is the effective attractive interaction.
To reproduce a spin-triplet odd-parity superconductivity, we employ a standard phenomenological attractive interaction~\cite{leggett_76, hara_86}:
\begin{align}
g_{\alpha \beta \gamma \delta}(\boldsymbol{k}, \boldsymbol{k}^{\prime})=
g_0 \left[ \Phi_x(\boldsymbol{k})\Phi_x(\boldsymbol{k}^{\prime})+\Phi_y(\boldsymbol{k})\Phi_y(\boldsymbol{k}^{\prime}) \right]
\label{eq:att_int}
\end{align}
for $|\xi_{\boldsymbol{k}}|, |\xi_{\boldsymbol{k}^{\prime}}| \leq \epsilon_c$, and
$g_{\alpha \beta \gamma \delta}(\boldsymbol{k}, \boldsymbol{k}^{\prime})=0$ for $|\xi_{\boldsymbol{k}}|, |\xi_{\boldsymbol{k}^{\prime}}| > \epsilon_c$,
where we assume that the attractive interaction acts only for the electrons having the kinetic energy in the range of $-\epsilon_c \leq \xi_{\boldsymbol{k}} \leq \epsilon_c$.
The pairing functions, $\Phi_{x}(\boldsymbol{k})=-\Phi_{x}(-\boldsymbol{k})$ and $\Phi_{y}(\boldsymbol{k})=-\Phi_{y}(-\boldsymbol{k})$,
have the same rotation properties as $k_x$ and $k_y$ under the $D_{4h}$ point group symmetry of Sr$_2$RuO$_4$.
In the bulk with the $D_{4h}$ point group symmetry, the spin-triplet superconducting states are classified into the four helical states
$\boldsymbol{d}_{\boldsymbol{k}}= (\Phi_{x}(\boldsymbol{k}) , \pm \Phi_{y}(\boldsymbol{k}), 0)$ and
$\boldsymbol{d}_{\boldsymbol{k}}= (\Phi_{y}(\boldsymbol{k}) , \pm \Phi_{x}(\boldsymbol{k}),0 )$,
and the two chiral states, $\boldsymbol{d}_{\boldsymbol{k}}= (0,0,\Phi_{y}(\boldsymbol{k}) \pm i \Phi_{x}(\boldsymbol{k}) )$.
In a real Sr$_2$RuO$_4$, the atomic spin-orbit coupling lifts the degeneracy of these six spin-triplet superconducting states.
In the single-band model, such effect is effectively reproduced by a spin-dependence of the attractive interaction~\cite{yanase_05}.
However, the previous theoretical~\cite{yanase_03} and experimental~\cite{murakawa_04, murakawa_07} studies suggest that
the lifting of degeneracy is very small in the bulk, where the splitting in transition temperature $T_c$ is estimated to be less than $0.01 T_c$~\cite{yanase_05}.
Therefore, we here ignore the spin-dependence of the attractive interaction.
Although the substantial form of $\Phi_{x(y)}(\boldsymbol{k})$ is still under discussion, on the basis of several microscopic theories~\cite{nomura_00, fujimoto_09, wang_13, simon_15},
we consider an odd-parity superconductivity within the next-nearest-neighbor pairing as
\begin{align}
\Phi_{x(y)}(\boldsymbol{k}) = \left[ \sin k_{x(y)} - \eta \cos k_{y(x)} \sin k_{x(y)} \right] / \Phi_0,
\label{eq:pair_func}
\end{align}
where the first term and second term represent the nearest-neighbor (NN) pairing and next-nearest-neighbor (NNN) pairing, respectively.
We normalize $\Phi_{x(y)}(\boldsymbol{k})$ by $\Phi_0$ so that the maximum value of $\Phi_{x(y)}(\boldsymbol{k})$ becomes unity.
Since several theories suggest that the NNN pairing becomes dominant~\cite{fujimoto_09, wang_13, simon_15, kusunose_05, kallin_15},
we consider both NN pairing dominant case ($\eta = 0.5$) and NNN pairing dominant case ($\eta=2.0$).
By substituting Eq.~(\ref{eq:att_int}) into Eq.~(\ref{eq:gap_eq}), we find that the pair potential is given in the form of
$\Delta_{\boldsymbol{k}, \alpha \beta} = \left[ \boldsymbol{d}_{\boldsymbol{k}} \cdot \hat{\boldsymbol{\sigma}} (i \hat{\sigma}_y) \right]_{\alpha \beta}$,
where the $\nu$ ($=x$,$y$,$z$) component of the $\boldsymbol{d}$-vector is represented by
\begin{align}
d_{\nu} (\boldsymbol{k}) = \Delta_0 \left[ X_{\nu} \Phi_{x}(\boldsymbol{k}) + Y_{\nu} \Phi_{y}(\boldsymbol{k}) \right],
\end{align}
with $X_{\nu}$ and $Y_{\nu}$ being the numerical coefficients determined by the gap equation in Eq.~(\ref{eq:gap_eq}).
To obtain $X_{\nu}$ and $Y_{\nu}$, we solve the gap equation by using an iterative method.
We show the detailed calculations for solving the gap equation in Supplemental Materials~\cite{supp_mat}.
In what follows, we focus on the gap function at zero-temperature.
The magnitude of attractive interaction $g_0$ is determined so that the amplitude of pair potential at $\boldsymbol{g}_{\boldsymbol{k}}=\boldsymbol{V}=0$ becomes $\Delta_0=0.001$.
The cutoff energy $\epsilon_c$ is fixed to $10 \Delta_0$.

\begin{figure}[tttt]
\begin{center}
\includegraphics[width=0.5\textwidth]{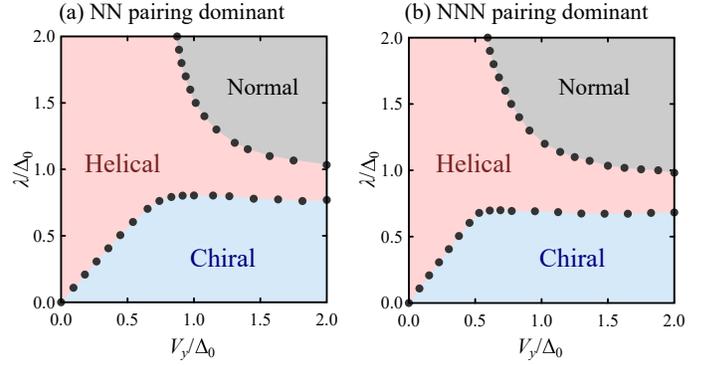}
\caption{Phase diagram as a function of the magnitude of ASOC potential $\lambda$ and Zeeman potential.
The magnetic field is applied to the $y$-direction as $\boldsymbol{V}=(0, V_y, 0)$.
In (a) and (b), we consider the NN pairing dominant case ($\eta = 0.5$) and NNN pairing dominant case ($\eta = 2.0$), respectively.}
\label{fig:figure1}
\end{center}
\end{figure}

In Figs.~\ref{fig:figure1}(a) and (b), we show the phase diagrams obtained from the gap equation
as a function of the magnitude of the ASOC potential $\lambda$ and Zeeman potential $|\boldsymbol{V}|$.
We consider the NN pairing dominant case ($\eta = 0.5$) and NNN pairing dominant case ($\eta = 2.0$) in Fig.~\ref{fig:figure1}(a) and (b), respectively.
We apply the magnetic field to the $y$-direction as $\boldsymbol{V}=(0, V_y, 0)$.
In both cases, for small Zeeman potentials, we obtain the helical state where the $\boldsymbol{d}$-vector is approximately given as
\begin{align}
\boldsymbol{d}^{h}_{\boldsymbol{k}} \propto (\Phi_{y}(\boldsymbol{k}) , - \delta \Phi_{x}(\boldsymbol{k}),0 ),
\end{align}
with $\delta$ being a real number in the range of $0<\delta<1$.
For small ASOC potentials and large Zeeman potentials, in both Figs.~\ref{fig:figure1}(a) and (b), we find the chiral state approximately described by
\begin{align}
\boldsymbol{d}^{c}_{\boldsymbol{k}} \propto (\Phi_{y}(\boldsymbol{k}) \pm i \Phi_{x}(\boldsymbol{k}),0 , 0),
\end{align}
where the $\boldsymbol{d}$-vector satisfies $\boldsymbol{d}^{c}_{\boldsymbol{k}} \perp \boldsymbol{V}$.
In contrast to the bulk chiral states, $\boldsymbol{d}^{c}_{\boldsymbol{k}}$ is pinned in the basal plane.
For the large ASOC and Zeeman potentials, we obtain the normal states with $\boldsymbol{d}_{\boldsymbol{k}}=0$.
The detailed structure of the $\boldsymbol{d}$-vector is shown in Supplemental Materials~\cite{supp_mat}.
At the phase boundary from the helical phase to the chiral phase,
the $\boldsymbol{d}$-vector suddenly changes from $\boldsymbol{d}^{h}_{\boldsymbol{k}}$ to $\boldsymbol{d}^{c}_{\boldsymbol{k}}$
(See also Supplemental Materials~\cite{supp_mat}).
This suggests that the helical-chiral phase transition is the first-order.
At the phase boundary from the helical phase to the normal phase, the amplitude of $\boldsymbol{d}^{h}_{\boldsymbol{k}}$ suddenly drops to zero.
This implies that the helical state undergoes a first-order phase transition to the normal state.
Even so, in analogy with spin-singlet superconductors~\cite{Fulde_64, Larkin_65, Yang_98},
there is a possibility that a Fulde-Ferrell-Larkin-Ovchinnikov (FFLO) state appears in the vicinity of the first-order phase boundary between the superconducting state and the normal state.
However, in this Rapid Communication, we focus on the phase transition from the helical state to the chiral state,
and the possibility for the FFLO phase in the vicinity of the normal phase is beyond the scope of this work. 
Importantly, the presence of the helical-chiral phase transition is well understood by the following two generic features of the $\boldsymbol{d}$-vector:
\begin{itemize}
\item[(i)] The pair-breaking effect of an ASOC potential $\boldsymbol{g}$ damages the component of the $\boldsymbol{d}$-vector perpendicular to $\boldsymbol{g}$~\cite{Frigeri_04}.
Thus, helical states are energetically favorable in the presence of the ASOC potential
because they can optimize the condensation energy from the relation of $\boldsymbol{d} \parallel \boldsymbol{g}$~\cite{yanase_13, fujimoto_09, Frigeri_04, Scheurer_19}.
\item[(ii)] The paramagnetic pair-breaking effect of a Zeeman potential $\boldsymbol{V}$ damages the component of the $\boldsymbol{d}$-vector parallel to $\boldsymbol{V}$.
Thus, the chiral states satisfying $\boldsymbol{d} \perp \boldsymbol{V}$ are energetically favorable in the presence of the Zeeman potential
because they are completely free from the paramagnetic pair-breaking effect~\cite{kusunose_05, yanase_14, oda_19}.
\end{itemize}
\noindent
The realization of the helical state of $\boldsymbol{d}^{h}_{\boldsymbol{k}}$ for small Zeeman potentials are mainly explained by the feature (i).
Namely, the helical state appears for small Zeeman potentials to minimize the dominant pair breaking effect from the ASOC potential.
The suppression in the $x$-component of $\boldsymbol{d}^{h}_{\boldsymbol{k}}$ characterized by $\delta$ is due to the paramagnetic pair breaking effect discussed in the feature (ii).
The realization of the chiral states of $\boldsymbol{d}^{c}_{\boldsymbol{k}}$ for large Zeeman potentials are naturally understood from the feature (ii).
The features (i) and (ii) are the rigid concepts irrelevant to the details of model.
Actually, it has been confirmed that the feature (i)~\cite{yanase_13, fujimoto_09,  Scheurer_19} and
feature (ii)~\cite{yanase_14, oda_19} are valid even with the multi-band models for the Sr$_2$RuO$_4$.
We also confirm that the helical-chiral phase transition occurs even when we employ a more realistic ASOC potential discussed in Ref.~[\onlinecite{yanase_13}].
Moreover, in principle, the amplitude of ASOC potentials can be tuned by changing the substrate, by fabricating capping layers, or by applying gate voltages.
Therefore, we can highly expect that spin-triplet superconducting thin-films show the helical-chiral phase transition in experiments.

\textit{Signature in tunneling spectroscopy}---%
Next, to approach the detection of the helical-chiral phase transition in experiments,
we study the differential conductance in a two-dimensional normal-metal/spin-triplet superconductor (NS) junction.
We assume that the NS junction consists of the semi-infinite normal-metal segment located for $x < x_0$
and the semi-infinite spin-triplet superconducting segment located for $x \geq x_0$,
where the periodic boundary condition is applied to the direction parallel to the junction interface (i.e., the $y$-direction).
We describe the normal segment by setting $\Delta_0=0$.
To describe the superconducting segment, we use the $\boldsymbol{d}$-vector obtained from the gap equation in Eq.~(\ref{eq:gap_eq}).
The hopping integrals between the normal and superconducting segments are chosen as $t=0.05$ and $t^{\prime}=0.0$ to describe a low transparency junction.
The Hamiltonian used for calculating the differential conductance is explicitly shown in Supplemental Materials~\cite{supp_mat}.
We calculate the differential conductance $G_{\mathrm{NS}}$ based on the formula~\cite{klapwijk_82, bruder_90, kashiwaya_00}
\begin{align}
G_{\mathrm{NS}}(eV) = \frac{e^{2}}{h} \sum_{\zeta,\zeta^{\prime}}
\left[ \delta_{\zeta,\zeta^{\prime}} - \left| r^{ee}_{\zeta,\zeta^{\prime}} \right|^{2}
+ \left| r^{he}_{\zeta,\zeta^{\prime}} \right|^{2} \right]_{eV=E},
\end{align}
where $r^{ee}_{\zeta,\zeta^{\prime}}$ and $r^{he}_{\zeta,\zeta^{\prime}}$ denote the normal and Andreev reflection coefficients at the energy $E$, respectively.
The index $\zeta$ and $\zeta^{\prime}$ label the outgoing and incoming channel, respectively.
These reflection coefficients are calculated by using the lattice Green's function techniques~\cite{fisher_81, ando_91}.
The results are normalized by the normal conductance $G_{\mathrm{N}}$, which is calculated by setting $\boldsymbol{V}=\boldsymbol{d}_{\boldsymbol{k}}=0$.

\begin{figure}[tttt]
\begin{center}
\includegraphics[width=0.5\textwidth]{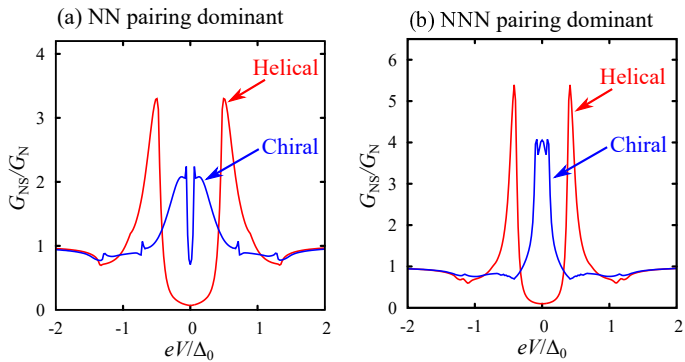}
\caption{Differential conductance as a function of the bias-voltage with (a)$\eta = 0.5$ and (b)$\eta = 2.0$.
We choose $V_y$ very close to the phase boundary: $V_y=V_c \pm 0.001 \Delta_0$ for the helical (chiral) state
with $V_c$ representing the critical magnitude of Zeeman potential at $\lambda=0.5 \Delta_0$.}
\label{fig:figure2}
\end{center}
\end{figure}

\begin{figure}[tttt]
\begin{center}
\includegraphics[width=0.5\textwidth]{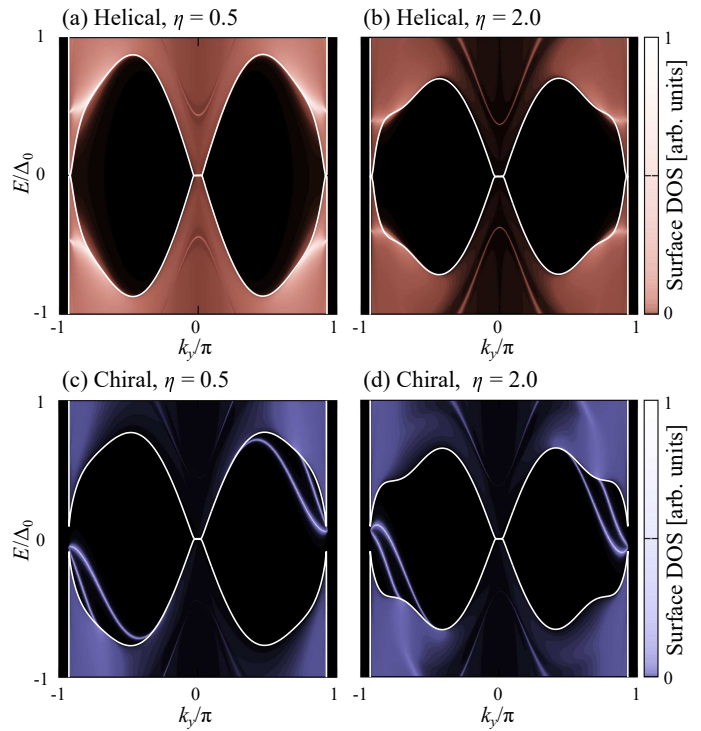}
\caption{Surface density of states for the semi-finite superconductor as a function of the energy and momentum parallel to the surface $k_y$.
For the helical [chiral] phases, shown in (a) and (b) [(c) and (d)], we use $V_y=V_c \pm 0.001 \Delta_0$.
The white lines denote the lowest bulk-energy dispersion obtained by diagonalizing the bulk Hamiltonian.}
\label{fig:figure3}
\end{center}
\end{figure}

In Figs.~\ref{fig:figure2}(a) and (b), we show the differential conductance of the NS junction
as a function of the bias voltage $eV$ with $\lambda=0.5\Delta_0$ and $\boldsymbol{V}=(0, V_y, 0)$.
We consider the NN pairing dominant case ($\eta = 0.5$) and NNN pairing dominant case ($\eta = 2.0$) in Fig.~\ref{fig:figure2}(a) and (b), respectively.
With $\lambda=0.5 \Delta_0$, the phase boundary between the helical and chiral phase is located at $V_y=0.441 \Delta_0$ ($0.373\Delta_0$) for $\eta=0.5$ ($\eta=2.0$).
We choose $V_y$ very close to the phase boundary: $V_y=V_c \pm 0.001 \Delta_0$ for the helical (chiral) state.
As shown in Figs.~\ref{fig:figure2}(a) and (b), the conductance spectra for the helical states (red line) show the U-shaped structures~\cite{sngupta_02}.
Nevertheless, when the helical states undergo the phase transition to the chiral states, the conductance spectra show the sudden enhancement in low bias voltages (blue line),
whereas the steep zero-bias dip is found for $\eta=0.5$.
The sudden change in the tunneling conductance implies that the properties of surface Andreev bound states (ABSs) are changed drastically through the phase transition.
To confirm this statement, we also calculate the surface density of states (DOS) for the semi-infinite superconductor
by using the formula $\rho_{k_y}(E)=-\mathrm{Im}[\mathrm{Tr}\; G_{k_y}(x_0,x_0,E+i\gamma)]/\pi$
with $G_{k_y}(x,x^{\prime},E)$ and $\gamma$ representing the Green's function and small imaginary part added to the energy, respectively.
$\mathrm{Tr}$ means the trace for spin and Nambu space of the Green's function.
To calculate the surface DOS, we remove the semi-infinite normal segment located for $x<x_0$.
The small imaginary part of the energy in the Green's function is chosen as $\gamma = 10^{-3} \Delta_0$.
In Figs.~\ref{fig:figure3}(a)-(d), we show the surface DOS as a function of the energy and momentum parallel to the surface $k_y$.
The white lines denote the lowest bulk energy for each $k_y$, which is obtained by diagonalizing the bulk Hamiltonian.
Due to the pair breaking effect of the ASOC and Zeeman potentials, the bulk superconducting gap vanishes partially in momentum space.
For the helical phases, as shown in Figs.~\ref{fig:figure3}(a) and (b), the ABSs are absent in low energies.
In contrast, as shown in Figs.~\ref{fig:figure3}(c) and (d), we can still find the inner-gap ABSs in the chiral phases.
This qualitative difference in the ABSs of the helical phase and that of the chiral phase is related with a generic concept regarding the topological classification~\cite{schnyder_08}:
\begin{itemize}
\item[(iii)] A helical (chiral) superconductor in two-dimensions can exhibit the surface ABSs characterized by a $\mathbb{Z}_2$ ($\mathbb{Z}$) topological invariant.
Thus, the ABSs of the helical (chiral) superconductor are intrinsically fragile (robust) against Zeeman potentials breaking time-reversal symmetry.
\end{itemize}
Strictly speaking, we can no longer employ the topological invariant for the present junction because the bulk superconducting gap is closed.
Even so, the absence (presence) of the ABSs in the helical (chiral) phase is well understood by their intrinsic fragility (robustness) against the Zeeman potential.
The detailed structure of the conductance spectra depends on the details of model for the Sr$_2$RuO$_4$ superconductors
~\cite{sngupta_02, yamashiro_97, honerkamp_98, wu_10, yada_14}.
However, according to the topological concepts (iii) irrelevant to the details of model,
the surface ABSs in the helical phase and that in the chiral phase have the distinctively different characters in the presence of the Zeeman potential.
Therefore, even in real experiments, we can highly expect that the helical-chiral phase transition can be detected through the drastic change in the tunneling conductance spectrum.

\textit{Summary}---%
In summary, we demonstrate that spin-triplet superconducting thin-films show the helical-chiral phase transition by applying in-plane magnetic fields (see Fig.~\ref{fig:figure1}).
This phase transition is unique in the presence of the $\boldsymbol{d}$-vector, and is intrinsically absent in spin-singlet superconductors.
The helical-chiral phase transition can be detected by the sudden change in the conductance spectrum of the NS junction (see Fig.~\ref{fig:figure2})
reflecting the drastic change in the properties of the surface ABSs through the phase transition (see Fig.~\ref{fig:figure3}).
Our proposal is constructed by the combination of the three fundamental and rigid concepts (i)-(iii) irrelevant to the details of the model.
Consequently, we propose a promising strategy for identifying the realization of spin-triplet superconductivity in Sr$_2$RuO$_4$ thin-films.

\begin{acknowledgments}
We are grateful to S. Yonezawa for fruitful discussions.
Y. T., S. K. and Y. A. was supported by Grants-in-Aid from JSPS for Scientific Research on Innovative Areas ``Topological Materials Science''
(KAKENHI Grant Numbers JP15H05851, JP15H05853, JP15H05855, and JP15K21717).
Y. T. is also supported by Scientific Research (B) (KAKENHI Grants No. JP18H01176).
\end{acknowledgments}

\clearpage

\pagebreak
\onecolumngrid
\begin{center}
  \textbf{\large Supplemental Materials for \\ ``Helical-Chiral Phase Transition in Sr$_2$RuO$_4$ Thin-Films \\ and its Detection via Tunneling Spectroscopy''}\\ \vspace{0.3cm}
S. Ikegaya$^{1}$, K. Yada$^{2}$, Y. Tanaka$^{2}$, S. Kashiwaya$^{2}$, Y. Asano$^{3}$, D. Manske$^{1}$\\ \vspace{0.1cm}
{\itshape $^{1}$Max-Planck-Institut f\"ur Festk\"orperforschung, Heisenbergstrasse 1, D-70569 Stuttgart, Germany\\
$^{2}$Department of Applied Physics, Nagoya University, Nagoya 464-8603, Japan\\
$^{3}$Department of Applied Physics, Hokkaido University, Sapporo 060-8628, Japan}
\date{\today}
\end{center}

\section{Detailed Calculations for Solving Gap Equation}
In this section, we show the detailed calculations for solving the gap equation in Eq.~(2) in the main text.
We here rewrite the mean-field Hamiltonian in Eq.~(1) in the main text as follows,
\begin{gather}
H=\frac{1}{2}\sum_{\boldsymbol{k}} \boldsymbol{C}^{\dagger}_{\boldsymbol{k}} \check{H}_{\boldsymbol{k}} \boldsymbol{C}_{\boldsymbol{k}}
+\frac{1}{2}\sum_{\boldsymbol{k},\alpha}\xi_{\boldsymbol{k}}
-\frac{1}{2}\sum_{\boldsymbol{k}}\sum_{\alpha,\beta}
\Delta_{\boldsymbol{k},\alpha \beta} \langle c^{\dagger}_{\boldsymbol{k} \alpha} c^{\dagger}_{-\boldsymbol{k} \beta} \rangle, \label{eq:original_ham}\\
\boldsymbol{C}_{\boldsymbol{k}}=
\left[ c_{\boldsymbol{k}\uparrow}, c_{\boldsymbol{k}\downarrow}, c^{\dagger}_{-\boldsymbol{k}\uparrow}, c^{\dagger}_{-\boldsymbol{k}\downarrow}
\right]^{\mathrm{T}},\\
\check{H}_{\boldsymbol{k}}=\left[ \begin{array}{cc}
\hat{h}_{\boldsymbol{k}} & \hat{\Delta}_{\boldsymbol{k}} \\
\hat{\Delta}^{\ast}_{\boldsymbol{k}} & -\hat{h}^{\ast}_{-\boldsymbol{k}}\\ \end{array} \right],\\
\hat{h}_{\boldsymbol{k}}=
\xi_{\boldsymbol{k}}\hat{\sigma}_0 + \boldsymbol{g}_{\boldsymbol{k}}\cdot\hat{\boldsymbol{\sigma}} + \boldsymbol{V}\cdot\hat{\boldsymbol{\sigma}}, \label{eq:norm_ham}\\
\hat{\Delta}_{\boldsymbol{k}}=\left[ \begin{array}{cc}
\Delta_{\boldsymbol{k},\uparrow \uparrow} & \Delta_{\boldsymbol{k},\uparrow \downarrow}\\
\Delta_{\boldsymbol{k},\downarrow \uparrow} & \Delta_{\boldsymbol{k},\downarrow \downarrow}\\
\end{array}\right]
=\left[ \begin{array}{cc}
-d_x (\boldsymbol{k})+i d_y (\boldsymbol{k}) & d_z(\boldsymbol{k}) \\
d_z(\boldsymbol{k}) & d_x (\boldsymbol{k})+i d_y (\boldsymbol{k}) \\
\end{array}\right], \label{eq:pair_potential} \\
d_{\nu} (\boldsymbol{k}) = \Delta_0 \left[ X_{\nu} \Phi_{x}(\boldsymbol{k}) + Y_{\nu} \Phi_{y}(\boldsymbol{k}) \right], \label{eq:d-vector}\\
\Phi_{x(y)}(\boldsymbol{k}) = \left[ \sin k_{x(y)} - \eta \cos k_{y(x)} \sin k_{x(y)} \right] / \Phi_0,  \label{eq:pair_function}
\end{gather}
where the numerical coefficient in $\boldsymbol{d}$-vector, $X_{\nu}$ and $Y_{\nu}$ for $\nu=x$, $y$, $z$, are determined by the gap equation,
\begin{gather}
\Delta_{\boldsymbol{k},\alpha\beta}= \sum_{\boldsymbol{k}^{\prime}} \sum_{\gamma, \delta}
g_{\alpha \beta \gamma \delta}(\boldsymbol{k}, \boldsymbol{k}^{\prime})
\langle c_{-\boldsymbol{k}^{\prime}\gamma} c_{\boldsymbol{k}^{\prime}\delta} \rangle, \label{eq:gap_equation} \\
g_{\alpha \beta \gamma \delta}(\boldsymbol{k}, \boldsymbol{k}^{\prime})=g(\boldsymbol{k}, \boldsymbol{k}^{\prime})=\left\{ \begin{array}{cl}
g_0 \left[ \Phi_x(\boldsymbol{k})\Phi_x(\boldsymbol{k}^{\prime})+\Phi_y(\boldsymbol{k})\Phi_y(\boldsymbol{k}^{\prime}) \right]
& \qquad \text{for } - \epsilon_c \leq \xi_{\boldsymbol{k}},\xi_{\boldsymbol{k}^{\prime}} \leq \epsilon_c,\\
0 & \text{otherwise}.
\end{array} \right. \label{eq:att_int}
\end{gather}
From Eqs.~(\ref{eq:pair_potential}), (\ref{eq:gap_equation}) and (\ref{eq:att_int}),
we obtain the gap equations for each component of the $\boldsymbol{d}$-vector as
\begin{align}
d_x(\boldsymbol{k})&=-\frac{\Delta_{\boldsymbol{k},\uparrow \uparrow}-\Delta_{\boldsymbol{k},\downarrow \downarrow}}{2} \nonumber\\
&=-\frac{1}{2}\sum_{\boldsymbol{k}^{\prime}}
g(\boldsymbol{k}, \boldsymbol{k}^{\prime}) \left[ \langle c_{-\boldsymbol{k}^{\prime} \uparrow} c_{\boldsymbol{k}^{\prime} \uparrow} \rangle
- \langle c_{-\boldsymbol{k}^{\prime} \downarrow} c_{\boldsymbol{k}^{\prime} \downarrow} \rangle \right] \nonumber\\
&=\frac{1}{2}\sum_{\boldsymbol{k}^{\prime}}
g(\boldsymbol{k}, \boldsymbol{k}^{\prime}) \left[ \langle c_{\boldsymbol{k}^{\prime} \uparrow} c_{-\boldsymbol{k}^{\prime} \uparrow} \rangle
- \langle c_{\boldsymbol{k}^{\prime} \downarrow} c_{-\boldsymbol{k}^{\prime} \downarrow} \rangle \right],\label{eq:dx}
\end{align}
\begin{align}
d_y(\boldsymbol{k})&=-i\frac{\Delta_{\boldsymbol{k},\uparrow \uparrow}+\Delta_{\boldsymbol{k},\downarrow \downarrow}}{2} \nonumber\\
&=-i\frac{1}{2}\sum_{\boldsymbol{k}^{\prime}}
g_{\boldsymbol{k},\boldsymbol{k}^{\prime}} \left[ \langle c_{-\boldsymbol{k}^{\prime} \uparrow} c_{\boldsymbol{k}^{\prime} \uparrow} \rangle
+ \langle c_{-\boldsymbol{k}^{\prime} \downarrow} c_{\boldsymbol{k}^{\prime} \downarrow} \rangle \right] \nonumber\\
&=i\frac{1}{2}\sum_{\boldsymbol{k}^{\prime}}
g_{\boldsymbol{k},\boldsymbol{k}^{\prime}} \left[ \langle c_{\boldsymbol{k}^{\prime} \uparrow} c_{-\boldsymbol{k}^{\prime} \uparrow} \rangle
+ \langle c_{\boldsymbol{k}^{\prime} \downarrow} c_{-\boldsymbol{k}^{\prime} \downarrow} \rangle \right],\label{eq:dy}
\end{align}
\begin{align}
d_z(\boldsymbol{k})&=\frac{\Delta_{\boldsymbol{k},\uparrow \downarrow}+\Delta_{\boldsymbol{k},\downarrow \uparrow}}{2} \nonumber\\
&=\frac{1}{2}\sum_{\boldsymbol{k}^{\prime}}
g_{\boldsymbol{k},\boldsymbol{k}^{\prime}} \left[ \langle c_{-\boldsymbol{k}^{\prime} \uparrow} c_{\boldsymbol{k}^{\prime} \downarrow} \rangle
+ \langle c_{-\boldsymbol{k}^{\prime} \downarrow} c_{\boldsymbol{k}^{\prime} \uparrow} \rangle \right] \nonumber\\
&=-\frac{1}{2}\sum_{\boldsymbol{k}^{\prime}}
g_{\boldsymbol{k},\boldsymbol{k}^{\prime}} \left[ \langle c_{\boldsymbol{k}^{\prime} \uparrow} c_{-\boldsymbol{k}^{\prime} \downarrow} \rangle
+ \langle c_{\boldsymbol{k}^{\prime} \downarrow} c_{-\boldsymbol{k}^{\prime} \uparrow} \rangle \right],\label{eq:dz}
\end{align}
where we use $g_{\boldsymbol{k},\boldsymbol{k}^{\prime}}=-g_{\boldsymbol{k},-\boldsymbol{k}^{\prime}}$
to obtain the third lines of Eqs.~(\ref{eq:dx})-(\ref{eq:dz}) for later convenience.
By substituting the explicit form of the attractive interaction in Eq.~(\ref{eq:att_int}) into the Eqs.~(\ref{eq:dx})-(\ref{eq:dz}),
we obtain the gap equations for each coefficient in the $\boldsymbol{d}$-vector (i.e., $X_{\nu}$ and $Y_{\nu}$) as
\begin{align}
X_x&=\frac{g_0}{2}{\sum_{\boldsymbol{k}}}^{\prime} \Phi_{x}(\boldsymbol{k})
\left[ \langle c_{\boldsymbol{k} \uparrow} c_{-\boldsymbol{k} \uparrow} \rangle
- \langle c_{\boldsymbol{k} \downarrow} c_{-\boldsymbol{k} \downarrow} \rangle \right],\label{eq:xx}\\
Y_x&=\frac{g_0}{2}{\sum_{\boldsymbol{k}}}^{\prime} \Phi_{y}(\boldsymbol{k})
\left[ \langle c_{\boldsymbol{k} \uparrow} c_{-\boldsymbol{k} \uparrow} \rangle
- \langle c_{\boldsymbol{k} \downarrow} c_{-\boldsymbol{k} \downarrow} \rangle \right],\\
X_y&=i\frac{g_0}{2}{\sum_{\boldsymbol{k}}}^{\prime} \Phi_{x}(\boldsymbol{k})
\left[ \langle c_{\boldsymbol{k} \uparrow} c_{-\boldsymbol{k} \uparrow} \rangle
+ \langle c_{\boldsymbol{k} \downarrow} c_{-\boldsymbol{k} \downarrow} \rangle \right],\\
Y_y&=i\frac{g_0}{2}{\sum_{\boldsymbol{k}}}^{\prime} \Phi_{y}(\boldsymbol{k})
\left[ \langle c_{\boldsymbol{k} \uparrow} c_{-\boldsymbol{k} \uparrow} \rangle
+ \langle c_{\boldsymbol{k} \downarrow} c_{-\boldsymbol{k} \downarrow} \rangle \right],\\
X_z&=-\frac{g_0}{2}{\sum_{\boldsymbol{k}}}^{\prime} \Phi_{x}(\boldsymbol{k})
\left[ \langle c_{\boldsymbol{k} \uparrow} c_{-\boldsymbol{k} \downarrow} \rangle
+ \langle c_{\boldsymbol{k} \downarrow} c_{-\boldsymbol{k} \uparrow} \rangle \right],\\
Y_z&=-\frac{g_0}{2}{\sum_{\boldsymbol{k}}}^{\prime} \Phi_{y}(\boldsymbol{k})
\left[ \langle c_{\boldsymbol{k} \uparrow} c_{-\boldsymbol{k} \downarrow} \rangle
+ \langle c_{\boldsymbol{k} \downarrow} c_{-\boldsymbol{k} \uparrow} \rangle \right],\label{eq:yz}
\end{align}
where $\sum^{\prime}_{\boldsymbol{k}}$ represent the summation over $\boldsymbol{k}$ satisfying $-\epsilon_c \leq \xi_{\boldsymbol{k}} \leq \epsilon_c$.
We now discuss how to calculate
$\langle c_{\boldsymbol{k} \alpha} c_{-\boldsymbol{k} \beta} \rangle$ and $\langle c^{\dagger}_{-\boldsymbol{k} \alpha} c^{\dagger}_{\boldsymbol{k} \beta} \rangle$ numerically.
By diagonalizing the Bogoliubov-de Gennes (BdG) Hamiltonian $\check{H}_{\boldsymbol{k}}$ numerically,
we obtain the matrix $\check{U}_{\boldsymbol{k}}$ satisfying
\begin{gather}
\check{U}^{\dagger}_{\boldsymbol{k}}\check{H}_{\boldsymbol{k}}\check{U}_{\boldsymbol{k}}=\check{E}_{\boldsymbol{k}},\quad
\check{E}_{\boldsymbol{k}}=\mathrm{diag} \left[E_{\boldsymbol{k},1},E_{\boldsymbol{k},2},E_{\boldsymbol{k},3},E_{\boldsymbol{k},4}\right].
\end{gather}
By using the Bogoliubov transformation defined by
\begin{gather}
\boldsymbol{C}_{\boldsymbol{k}}=\check{U}_{\boldsymbol{k}} \boldsymbol{\Gamma}_{\boldsymbol{k}},\qquad
\boldsymbol{\Gamma}_{\boldsymbol{k}}=
\left[ \gamma_{\boldsymbol{k},1}, \gamma_{\boldsymbol{k},2}, \gamma_{\boldsymbol{k},3}, \gamma_{\boldsymbol{k},4} \right]^{\mathrm{T}}
\label{eq:bdg_trs}
\end{gather}
with $\gamma_{\boldsymbol{k},j}$ for $j=1$-$4$ being the annihilation operator of the Bogoliubov quasi-particle satisfying
\begin{align}
\{\gamma_{\boldsymbol{k},j},\gamma^{\dagger}_{\boldsymbol{k}^{\prime},j^{\prime}} \}
=\delta_{\boldsymbol{k},\boldsymbol{k}^{\prime}}\delta_{j,j^{\prime}}, \quad
\{\gamma_{\boldsymbol{k},j},\gamma_{\boldsymbol{k}^{\prime},j^{\prime}} \}=0,
\end{align}
the mean-filed Hamiltonian $H$ in Eq.~(\ref{eq:original_ham}) is deformed as
\begin{align}
H=\frac{1}{2}\sum_{\boldsymbol{k}} \sum_{j=1\text{-}4} E_{\boldsymbol{k},j}\gamma^{\dagger}_{\boldsymbol{k},j}\gamma_{\boldsymbol{k},j}
+\frac{1}{2}\sum_{\boldsymbol{k},\alpha}\xi_{\boldsymbol{k}}-\frac{1}{2}\sum_{\boldsymbol{k}}\sum_{\alpha,\beta}
\Delta_{\boldsymbol{k},\alpha \beta} \langle c^{\dagger}_{\boldsymbol{k} \alpha} c^{\dagger}_{-\boldsymbol{k} \beta} \rangle. \label{eq:bdg_diagonal}
\end{align}
To obtain $\langle c_{\boldsymbol{k} \alpha} c_{-\boldsymbol{k} \beta} \rangle$ and $\langle c^{\dagger}_{-\boldsymbol{k} \alpha} c^{\dagger}_{\boldsymbol{k} \beta} \rangle$,
we consider
\begin{gather}
\langle \boldsymbol{C}_{\boldsymbol{k}} \boldsymbol{C}^{\dagger}_{\boldsymbol{k}} \rangle=
\left[ \begin{array}{cc} \hat{A}_{\boldsymbol{k}} & \hat{B}_{\boldsymbol{k}} \\
\underline{\hat{B}}_{\boldsymbol{k}} & \underline{\hat{A}}_{\boldsymbol{k}} \\ \end{array} \right], \label{eq:cc1}\\
\left( \hat{A}_{\boldsymbol{k}}\right)_{\alpha,\beta}=\langle c_{\boldsymbol{k} \alpha} c^{\dagger}_{\boldsymbol{k} \beta} \rangle, \quad
\left( \underline{\hat{A}}_{\boldsymbol{k}}\right)_{\alpha,\beta}=\langle c^{\dagger}_{-\boldsymbol{k} \alpha} c_{-\boldsymbol{k} \beta} \rangle,\\
\left( \hat{B}_{\boldsymbol{k}}\right)_{\alpha,\beta}=\langle c_{\boldsymbol{k} \alpha} c_{-\boldsymbol{k} \beta} \rangle, \quad
\left( \underline{\hat{B}}_{\boldsymbol{k}}\right)_{\alpha,\beta}=\langle c^{\dagger}_{-\boldsymbol{k} \alpha} c^{\dagger}_{\boldsymbol{k} \beta} \rangle.
\end{gather}
By using Eq.~(\ref{eq:bdg_trs}), we can deform Eq.~(\ref{eq:cc1}) as
\begin{align}
\langle \boldsymbol{C}_{\boldsymbol{k}} \boldsymbol{C}^{\dagger}_{\boldsymbol{k}} \rangle &=
\check{U}_{\boldsymbol{k}} \langle \boldsymbol{\Gamma}_{\boldsymbol{k}} \boldsymbol{\Gamma}_{\boldsymbol{k}}\rangle \check{U}^{\dagger}_{\boldsymbol{k}} \nonumber\\
&=\check{U}_{\boldsymbol{k}} \left[ \begin{array}{cccc}
\langle \gamma_{\boldsymbol{k}, 1} \gamma^{\dagger}_{\boldsymbol{k},1} \rangle & 0 & 0 & 0  \\
0 & \langle \gamma_{\boldsymbol{k}, 2} \gamma^{\dagger}_{\boldsymbol{k},2} \rangle & 0 & 0  \\
0 & 0 & \langle \gamma_{\boldsymbol{k}, 3} \gamma^{\dagger}_{\boldsymbol{k},3} \rangle & 0  \\
0 & 0 & 0 & \langle \gamma_{\boldsymbol{k}, 4} \gamma^{\dagger}_{\boldsymbol{k},4} \rangle  \\ \end{array}\right]  \check{U}^{\dagger}_{\boldsymbol{k}} \nonumber\\
&=\check{U}_{\boldsymbol{k}} \left[ \begin{array}{cccc}
1-f(E_{\boldsymbol{k}, 1})  & 0 & 0 & 0  \\
0 & 1-f(E_{\boldsymbol{k}, 2})  & 0 & 0  \\
0 & 0 & 1-f(E_{\boldsymbol{k}, 3})  & 0  \\
0 & 0 & 0 & 1-f(E_{\boldsymbol{k}, 4})  \\
\end{array}\right]  \check{U}^{\dagger}_{\boldsymbol{k}}, \label{eq:cc2}
\end{align}
where we use $\langle \gamma_{\boldsymbol{k}, j} \gamma^{\dagger}_{\boldsymbol{k},j^{\prime}} \rangle=\delta_{j,j^{\prime}} [ 1 - f(E_{\boldsymbol{k}, j}) ]$ with
\begin{align}
f(E_{\boldsymbol{k}, j})=\frac{1}{2} \left[ 1 - \tanh \left( \frac{E_{\boldsymbol{k}, j}}{2T} \right) \right]
\end{align}
being the Fermi distribution function and with $T$ representing the temperature.
Since $\check{U}_{\boldsymbol{k}}$ and $\check{E}_{\boldsymbol{k}}$ are obtained numerically, from Eq.~(\ref{eq:cc2}),
we can compute $\hat{B}_{\boldsymbol{k}}$ and $\underline{\hat{B}}_{\boldsymbol{k}}$ including
$\langle c_{\boldsymbol{k} \alpha} c_{-\boldsymbol{k} \beta} \rangle$ and $\langle c^{\dagger}_{-\boldsymbol{k} \alpha} c^{\dagger}_{\boldsymbol{k} \beta} \rangle$, respectively.
By using the numerically obtained $\langle c_{\boldsymbol{k} \alpha} c_{-\boldsymbol{k} \beta} \rangle$,
we can solve the gap equation in Eqs.~(\ref{eq:xx})-(\ref{eq:yz})  by means of an iterative method.
More specifically, we first prepare 30 sets of the initial $\boldsymbol{d}$-vectors for the iteration,
where the coefficients $X_{\nu}$ and $Y_{\nu}$ for each initial $\boldsymbol{d}$-vector are determined randomly.
From the iteration, we obtain the 30 solutions for the gap equation corresponding to the 30 initial $\boldsymbol{d}$-vectors.
We finally chose the most energetically stable solution having the largest condensation energy defined as
\begin{align}
E_{\mathrm{cond}}=-\left[\langle H \rangle - \langle H \rangle_{\Delta_0 = 0}\right],
\end{align}
where the second term represents the expectation value of the Hamiltonian in the absence of the pair potential.
The expectation value of the normal Hamiltonian $\langle H \rangle_{\Delta_0 = 0}$ is explicitly given by
\begin{gather}
\langle H \rangle_{\Delta_0 = 0}=\sum_{\boldsymbol{k}}\sum_{s=\pm}\varepsilon_{\boldsymbol{k},s} f(\varepsilon_{\boldsymbol{k},s}), \\
\varepsilon_{\boldsymbol{k},\pm}=\xi_{\boldsymbol{k}}\pm \sqrt{(V_x+\lambda \sin k_y)^2 +(V_y-\lambda \sin k_x )^2},
\end{gather}
where we obtain $\varepsilon_{\boldsymbol{k},\pm}$ by diagonalizing $\hat{h}_{\boldsymbol{k}}$ in Eq.~(\ref{eq:norm_ham}).
The expectation value of the mean-field Hamiltonian $\langle H \rangle$ is explicitly given as
\begin{align}
\langle H \rangle &=\frac{1}{2}\sum_{\boldsymbol{k}} \sum_{j=1\text{-}4}
E_{\boldsymbol{k},j} \langle \gamma^{\dagger}_{\boldsymbol{k},j}\gamma_{\boldsymbol{k},j} \rangle
+\frac{1}{2}\sum_{\boldsymbol{k},\alpha}\xi_{\boldsymbol{k}}-\frac{1}{2}\sum_{\boldsymbol{k}}\sum_{\alpha,\beta}
\Delta_{\boldsymbol{k},\alpha \beta} \langle c^{\dagger}_{\boldsymbol{k} \alpha} c^{\dagger}_{-\boldsymbol{k} \beta}\rangle \nonumber\\
&=\frac{1}{2}\sum_{\boldsymbol{k}} \sum_{j=1\text{-}4}
E_{\boldsymbol{k},j} f(E_{\boldsymbol{k}, j})
+\frac{1}{2}\sum_{\boldsymbol{k},\alpha}\xi_{\boldsymbol{k}}+\frac{1}{2}\sum_{\boldsymbol{k}}\sum_{\alpha,\beta}
\Delta_{\boldsymbol{k},\alpha \beta} \langle c^{\dagger}_{-\boldsymbol{k} \alpha} c^{\dagger}_{\boldsymbol{k} \beta}\rangle, \label{eq:econd_sc}
\end{align}
where we use $\Delta_{\boldsymbol{k},\alpha \beta}=-\Delta_{-\boldsymbol{k},\alpha \beta}$ to obtain the last line of Eq.~(\ref{eq:econd_sc}) for convenience.
By using the numerically obtained $\langle c^{\dagger}_{-\boldsymbol{k} \alpha} c^{\dagger}_{\boldsymbol{k} \beta} \rangle$,
we can compute the condensation energy $E_{\mathrm{cond}}=-\left[\langle H \rangle - \langle H \rangle_{\Delta_0 = 0}\right]$,
as well as $\langle H \rangle$ in Eq.~(\ref{eq:econd_sc}).

We note that the normal Hamiltonian $\hat{h}_{\boldsymbol{k}}$ satisfies the relation of
\begin{align}
\hat{R} \; \hat{h}_{\boldsymbol{k}} \; \hat{R}^{\dagger} = \hat{h}_{\boldsymbol{k}},
\quad \hat{R} = \hat{\sigma}_x {\cal K},
\end{align}
where ${\cal K}$ represents the complex conjugation operator.
Thus, the superconducting state with the pair potential $\hat{\Delta}_{\boldsymbol{k}}$ is always degenerate with the superconducting state with
$\hat{\Delta}^{\prime}_{\boldsymbol{k}}=\hat{R} \; \hat{\Delta}_{\boldsymbol{k}} \; \hat{R}^{\mathrm{T}}$.
Moreover, the superconducting state with $\hat{\Delta}_{\boldsymbol{k}}$
is degenerate with the superconducting states with $ e^{i\phi} \hat{\Delta}_{\boldsymbol{k}}$, where $\phi$ represents the arbitrary $U(1)$ gauge.
In our numerical calculation, we choose one from these degenerate solutions:
we fix the $U(1)$ gauge satisfying $\mathrm{Im}[X_x]=0$, and find the solution satisfying $\mathrm{Im}[Y_x] \geq 0$.

\clearpage

\section{Detailed Structure of the $\boldsymbol{d}$-vector}
\begin{figure}[hhhh]
\begin{center}
\includegraphics[width=0.9\textwidth]{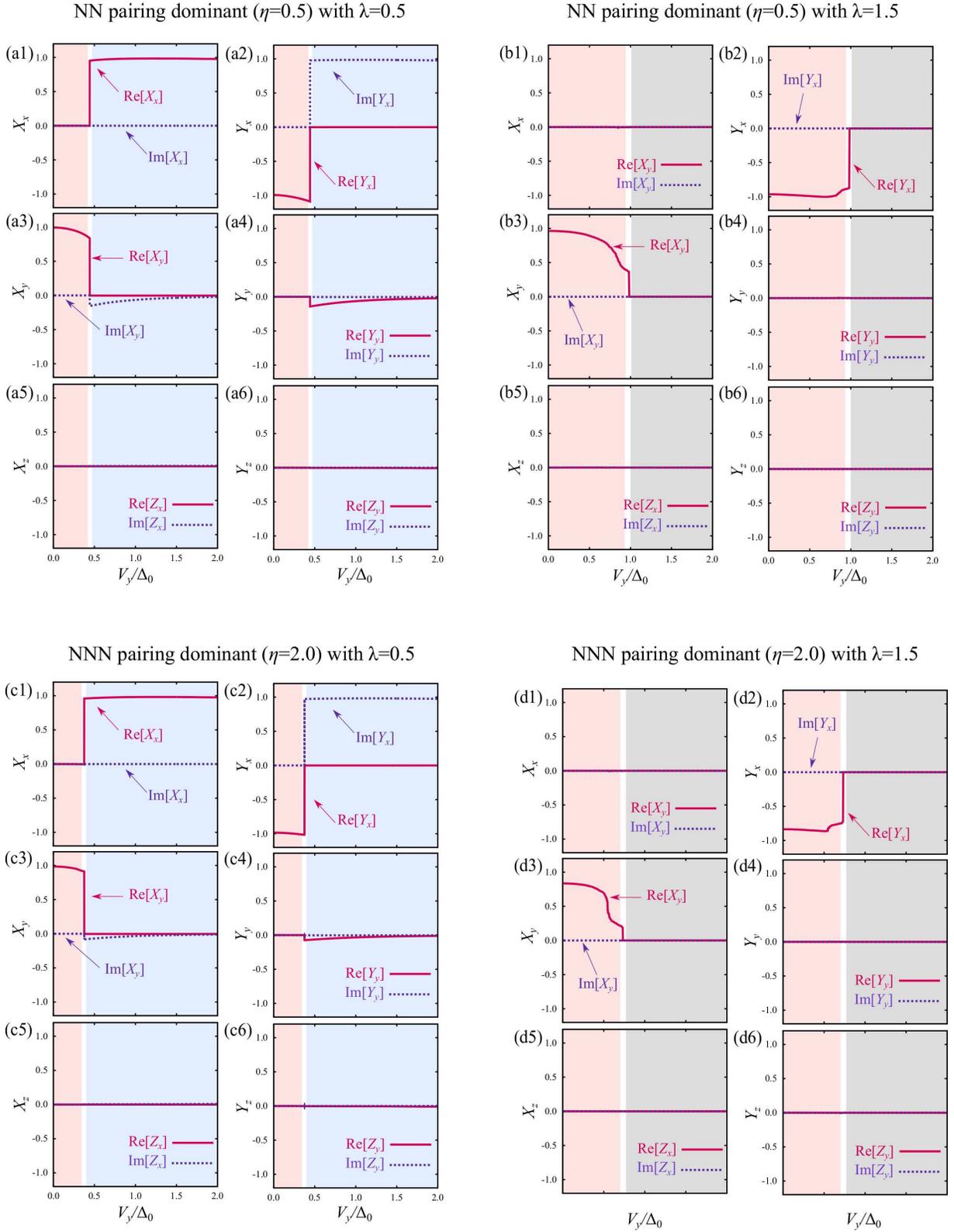}
\caption{Coefficients of the $\boldsymbol{d}$-vector, $X_{\nu}$ and $Y_{\nu}$, as a function of the Zeeman potential $V_y$.
In (a1)-(a6) and (b1)-(b6), we consider the NN pairing dominant case ($\eta=0.5$) with $\lambda=0.5\Delta_0$ and $\lambda=1.5\Delta_0$, respectively.
In (c1)-(c6) and (d1)-(d6), we consider the NNN pairing dominant case ($\eta=1.5$) with $\lambda=0.5\Delta_0$ and $\lambda=1.5\Delta_0$, respectively.
}
\label{fig:supplement1}
\end{center}
\end{figure}

In this section, we show the detailed structure of the $\boldsymbol{d}$-vector obtained from the gap equation in Eq.~(\ref{eq:gap_equation}).
In Figs.~\ref{fig:supplement1} (a1)-(a6), (b1)-(b6), (c1)-(c6), and (d1)-(d6),
we show the coefficients of the $\boldsymbol{d}$-vector, $X_{\nu}$ and $Y_{\nu}$, as a function of the Zeeman potential.
The magnetic field is applied to the $y$-direction as $\boldsymbol{V}=(0, V_y, 0)$.
In Figs.~\ref{fig:supplement1} (a1)-(a6), we show the results for the nearest-neighbor (NN) pairing dominant case ($\eta=0.5$)
with the antisymmetric spin-orbit coupling (ASOC) potential $\lambda=0.5 \Delta_0$.
For the Zeeman potentials smaller than a critical magnitude $V_c = 0.441 \Delta_0$,
we find that $X_y$ ($Y_x$) has the negative (positive) finite real value, while other coefficients is almost zero.
Moreover, the relation of $\mathrm{Re}[X_y]\geq\mathrm{Re}[Y_x]$ holds.
Thus, the $\boldsymbol{d}$-vector for $V_y<V_c$ is approximately given in the form of 
\begin{align}
\boldsymbol{d}_{\boldsymbol{k}} \propto (-\Phi_{y}(\boldsymbol{k}) , \delta \Phi_{x}(\boldsymbol{k}),0 ),
\end{align}
with $0<\delta<1$, which is equivalent to the $\boldsymbol{d}$-vector of the helical state discussed in the main text,
\begin{align}
\boldsymbol{d}^{h}_{\boldsymbol{k}} \propto (\Phi_{y}(\boldsymbol{k}) , -\delta \Phi_{x}(\boldsymbol{k}),0 ).
\end{align}
When the Zeeman potential across the critical value $V_c = 0.441 \Delta_0$,
the structure of the $\boldsymbol{d}$-vector is suddenly changed.
For $V_y>V_c$, we find that $X_x$ ($X_y$) becomes finite real (pure imaginary) number, while other coefficients are almost zero.
Moreover, the relation of $|X_x| \approx |X_y|$ holds.
Therefore, the $\boldsymbol{d}$-vector for $V_y>V_c$ is approximately given in the form of
\begin{align}
\boldsymbol{d}^{c}_{\boldsymbol{k}} \propto (\Phi_{y}(\boldsymbol{k}) \pm i \Phi_{x}(\boldsymbol{k}),0 , 0),
\end{align}
which describes the chiral states of the spin-triplet superconductor discussed in the main text.
In Figs.~\ref{fig:supplement1} (b1)-(b6), we show the results for the NN pairing dominant case ($\eta=0.5$) with $\lambda=1.5 \Delta_0$.
For small Zeeman potentials, we find the helical states described by $\boldsymbol{d}^{h}_{\boldsymbol{k}}$.
However, when the Zeeman potential exceeds the critical value of $V_c^{\prime}=0.981 \Delta_0$, all coefficients of the $\boldsymbol{d}$-vector suddenly drop to zero.
Namely, we obtain the normal states for $V_y>V^{\prime}_c$.
In Figs.~\ref{fig:supplement1} (c1)-(c6) and (d1)-(d6), we show the results for next-nearest-neighbor (NNN) pairing dominant case ($\eta=2.0$)
with $\lambda=0.5\Delta_0$ and $\lambda=1.5\Delta_0$, respectively.
Although the critical values for the phase transition are different, we obtain the qualitatively the same results as in the NN pairing dominant case.
Namely, even with the NNN pairing dominant case,
we find the phase transition from the helical state of $\boldsymbol{d}^{h}_{\boldsymbol{k}}$ to the chiral state of $\boldsymbol{d}^{c}_{\boldsymbol{k}}$
for the small ASOC potential ($\lambda=0.5 \Delta_0$),
and find the phase transition from the helical state to the normal state of $\boldsymbol{d}_{\boldsymbol{k}}=0$ with the large ASOC potential ($\lambda=1.5 \Delta_0$).
The phase diagrams in the main text is obtained by calculating the detailed structure of the $\boldsymbol{d}$-vector for various amplitudes of ASOC potentials $\lambda$.

\clearpage

\section{Hamiltonian for normal-metal/superconductor junctions}
\begin{figure}[hhhh]
\begin{center}
\includegraphics[width=0.5\textwidth]{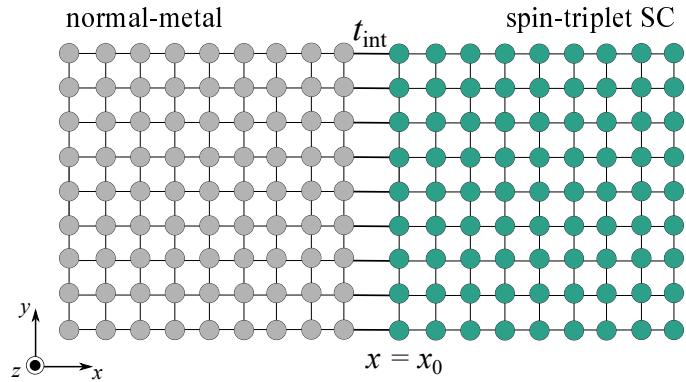}
\caption{Schematic image of the normal-metal/spin-triplet superconductor junction on the tight-binding model.
The junction interface is located at $x=x_0$, where the hopping integral between the normal-metal and superconductor segment is given by $t_{\mathrm{int}}$.
In the $y$-direction, we apply the periodic boundary condition.}
\label{fig:supplement2}
\end{center}
\end{figure}

In this section, we show the Hamiltonian used for calculating the differential conductance in a normal-metal/superconductor junction.
We consider the junction interface perpendicular to the $x$-direction,
where the superconductor (normal-metal) segment located for $x\geq x_0$ ($x < x_0$) as shown in Fig.~\ref{fig:supplement2}.
In the $y$-direction, the periodic boundary condition is applied.
We describe the present junction by the BdG Hamiltonian as
\begin{gather}
H_{\mathrm{NS}}=\frac{1}{2}\sum_{k_y}\sum_{x=-\infty}^{\infty} \left[
\boldsymbol{C}^{\dagger}_{x+1} \check{T}_x \boldsymbol{C}_{x}
+\boldsymbol{C}^{\dagger}_{x} \check{T}^{\dagger}_x \boldsymbol{C}_{x+1}
+\boldsymbol{C}^{\dagger}_{x} \check{H}_x \boldsymbol{C}_{x}\right],\\
\boldsymbol{C}_x=\left[ c_{x,k_y,\uparrow}, c_{x,k_y,\downarrow}, c^{\dagger}_{x,-k_y,\uparrow}, c^{\dagger}_{x,-k_y,\downarrow}\right]^{\mathrm{T}},\\
\check{T}_x=\left\{ \begin{array}{cl}
\check{T}_N & \text{for } x<x_0-1 \\
\check{T}_{\mathrm{int}} & \text{for } x=x_0-1 \\
\check{T}_S & \text{for } x \geq x_0
\end{array} \right. , \quad
\check{H}_j=\left\{ \begin{array}{cl}
\check{H}_N & \text{for } x < x_0 \\
\check{H}_S & \text{for } x \geq x_0
\end{array} \right. ,\\
\check{T}_N=\left[ \begin{array}{cc} \hat{t}_N(k_y) & 0 \\ 0 & -\hat{t}_N(k_y) \end{array} \right], \quad
\check{T}_S=\left[ \begin{array}{cc} \hat{t}_N(k_y) & \hat{t}_{\Delta} \\ -\hat{t}_{\Delta}^{\ast} & -\hat{t}_N(k_y) \end{array} \right], \quad
\check{T}_{\mathrm{int}}=\left[ \begin{array}{cc} -t_{\mathrm{int}}\hat{\sigma}_0 & 0 \\ 0 & t_{\mathrm{int}}\hat{\sigma}_0 \end{array} \right],\\
\check{H}_N=\left[ \begin{array}{cc} \hat{h}_N(k_y) & 0 \\ 0 & -\hat{h}^{\ast}_N(-k_y) \end{array} \right],\quad
\check{H}_S=\left[ \begin{array}{cc} \hat{h}_N(k_y) & \hat{h}_{\Delta}(k_y) \\ -\hat{h}^{\ast}_{\Delta}(-k_y) & -\hat{h}^{\ast}_N(-k_y) \end{array} \right],\\
\hat{t}_N=(-t-2t^{\prime} \cos k_y) \hat{\sigma}_0 - \frac{i \lambda}{2} \hat{\sigma}_y,\\
\hat{h}_N(k_y)=(-2 t \cos k_y - \mu) \hat{\sigma}_0+(\lambda \sin k_y + V_x)\hat{\sigma}_x+V_y \hat{\sigma}_y,\\
\hat{t}_{\Delta}=\frac{i \Delta_0}{2}\left[ -X_x \hat{\sigma}_z + i X_y \hat{\sigma}_0 + X_z \hat{\sigma}_x \right](1-\eta \cos k_y)
-\frac{\eta \Delta_0}{2}\left[ -Y_x \hat{\sigma}_z + i Y_y \hat{\sigma}_0 + Y_z \hat{\sigma}_x \right] \sin k_y,\\
\hat{h}_{\Delta}(k_y)=\left[ -Y_x \hat{\sigma}_z + i Y_y \hat{\sigma}_0 + Y_z \hat{\sigma}_x \right]\sin k_y,
\end{gather}
where $c^{\dagger}_{x,k_y,\alpha}$ ($c_{x,k_y,\alpha}$) is creation (annihilation) operator of an electron at $x$-th layer
with momentum parallel to the junction interface $k_y$ and spin $\alpha$.
The hopping integral at the junction interface is given by $t_{\mathrm{int}}$.
In the main text, we chose $t_{\mathrm{int}}=0.05$ to describe the low-transparency junction.
We determine the coefficients in the pair potential, $X_{\nu}$ and $Y_{\nu}$ for $\nu=x$, $y$, $z$, by the gap equation in Eq.~(\ref{eq:gap_equation}).

\end{document}